\documentstyle[12pt]{article}

\begin{document}
\makeatletter


\title{\bf  Strong CP problem, Neutron EDM and Thermal QCD sum rules}

\author {{Mohamed Chabab} \\ \\
{\it LPHEA, Physics  Department, Faculty of Science-Semlalia, }
\\ {\it Cadi Ayyad University, P.O. Box 2390, Marrakech, Morocco}
\\{\it mchabab@ucam.ac.ma} }
\date{}

\maketitle

\begin{abstract}
The behaviour of the broken CP symmetry at finite temperature is
examined. This is achieved through the investigation of the
neutron electric dipole moment $d_n$ induced by $\theta$-term. By
using thermal QCD sum rules, we find that below the critical
temperature,  the ratio $\mid {d_n \over \bar{\theta}}\mid$
slightly decreases but survives at temperature effects. This
evolution implies that CP remains broken at finite temperature as
required by Baryogenesis \cite{Chabab}.

\end{abstract}

\section{Introduction}

Recently the finite temperature behaviour of symmetries has gained
considerable interest. The question of symmetry restoration is a
non trivial phenomenon, since it has been shown in \cite{W,MS}
that more heat does not necessarily imply more symmetry. Besides,
the breaking of the symmetries has a profound implications in a
particle physics and cosmology. The CP symmetry is certainly one
of the most fundamental symmetries in nature. Besides its role in
solving domain wall problem \cite{ZKO}, it is a crucial
ingredient to understand the matter-antimatter asymmetry \cite{S}.

In the three generation Standard Model, CP violation originates
from the more obscure sector of the SM: the scalar part. It is
parameterized, in the electroweak sector, by a single phase
occuring in the Cabbibo-Kobayashi-Maskawa (CKM) quark mixing
matrix \cite{CKM}. CP violation could also originate from
additional CP-odd four dimensional operator embedded in the
following topological term  "$\theta$-term" in the QCD Lagrangian:

\begin{equation}
L_{\theta}=\theta  {\alpha_s\over 8
\pi}G_{\mu\nu}\tilde{G}^{\mu\nu},
\end{equation}

\quad which breaks P, T and CP. $G_{\mu\nu}$ is the gluonic field
strength, $\tilde{G}^{\mu\nu}$ denotes its dual and $\alpha_s$ is
the strong coupling constant. The  $G_{\mu\nu}\tilde{G}^{\mu\nu}$
quantity is a total derivative which contributes to the physical
observables only through non perturbative effects, induced by
instantons. A non zero value of $\theta$ may generate, in
particular, a sizable neutron electric dipole moment (NEDM) which
is related to the $\bar \theta$-angle by the following equation
obtained within the framework of Chiral perturbation theory:

\begin{equation}
d_n\sim {e\over M_n}({m_q\over M_n})\bar \theta \sim \{
\begin{array}{c}
2.7\times 10^{-16}\overline{\theta }\qquad \cite{Baluni}\\
5.2\times 10^{-16}\overline{\theta }\qquad \cite{cvvw}
\end{array}
\end{equation}
High precision Experiments have constrained the NEDM to
$d_n<1.1\times 10^{-25}ecm$ \cite{data}, providing a stringent
upper limit to $\bar \theta < 2\times10^{-10}$ \cite{peccei2}.
The difficulty to explain the smallness of $\bar{\theta}$ in the
standard model is usually known as the "strong CP problem". In
this regard, several scenarios were suggested. The most elegant
explanation is due to Peccei and Quinn \cite{PQ}, who identified
$\bar{\theta}$ to the axion, a very light pseudo scalar boson
arising from the spontaneous breaking of a global $U_A(1)$
symmetry. This particle may well be important to explain dark
matter puzzle providing a peace of information on the missing mass
of the universe \cite{LS}.

Our aim in this work is to investigate the behaviour of the CP
symmetry breaking at finite temperature and the thermal effects on
the restoration of the strong CP problem. This is motivated by the
possibility to restore some broken symmetries by increasing the
temperature.

In section 2, we perform the the calculations of the the $\bar
{\theta}$ induced NEDM using thermal QCD sum rules. Section 3 is
devoted to the discussion and qualitative analysis of the thermal
effects on the CP symmetry restoration.

\section{ NEDM from thermal QCD sum rules}

In order to derive the NEDM through the QCD sum rules techniques
\cite{SVZ,PR,PR1}, we consider a
 Lagrangian containing the following P and CP violating operators:
\begin{equation}
L_{P,CP}=-\theta_q m_* \sum_f \bar{q}_f i\gamma_5 q_f +\theta
{\alpha_s\over 8 \pi}G_{\mu\nu}\tilde{G}^{\mu\nu}.
\end{equation}
$\theta_q$ and $\theta$ are respectively two angles coming from
the chiral and the topological terms while $m_*$ is the quark
reduced mass given by $m_*$=$m_um_d \over{m_u +m_d} $. the
physical phase is $\bar{\theta}=\theta+\theta_q$. We usually
start from the two points correlators in QCD background with a
non-vanishing $\theta$ in the presence of a constant external
electomagnetic field $ F^{\mu\nu}$:

\begin{equation}
\Pi(q^2) = i \int d^4x
e^{iqx}<0|T\{\eta(x)\bar{\eta}(0)\}|0>_{\theta,F} .
\end{equation}
where $\eta(x)$ is the neutron  interpolating current  \cite{I}:
\begin{equation}
\eta
=2\epsilon_{abc}\{(d^T_aC\gamma_5u_b)d_c+\beta(d^T_aCu_b)\gamma_5d_c\},
\end{equation}
and  $\beta$ is a mixing parameter. To select the appropriate
Lorentz structure,
 $\Pi(q^2)$ is expanded in terms of the electromagnetic charge as:
\begin{equation}
\Pi(q^2)=\Pi^{(0)}(q^2)+e \Pi^{(1)}(q^2,F^{\mu\nu}) + O(e^2).
\end{equation}
The first term $\Pi^{(0)}(q^2)$ is the nucleon propagator which
includes only the CP-even parameters, while the second term
$\Pi^{(1)}(q^2,F^{\mu\nu})$ is the polarization tensor which may
be expanded through Wilson OPE as: $\sum C_n<0|\bar{q}\Gamma
q|0>_{\theta,F}$, where $\Gamma$ is an arbitrary Lorentz
structure and $C_n$ are the Wilson coefficient functions
calculable in  perturbation theory  \cite{SVZ1,IS}. From this
expansion, we keep  only the CP-odd contribution part. The
electromagnetic dependence of these matrix elements is determined
in terms of the magnetic  susceptibilities  $\kappa$, $\chi $ and
$\xi$ \cite{IS},  defined as:
\begin{eqnarray}
<0|\bar{q}\sigma^{\mu\nu} q|0>_F&=& \chi e_q F^{\mu\nu}
<0|\bar{q}q|0>
\\
g<0|\bar{q}G^{\mu\nu} q|0>_{F}&=& \kappa e_q F^{\mu\nu} <0|\bar{q}q|0> \\
2g<0|\bar{q}\tilde{G}^{\mu\nu} q|0>_{F}&=& \xi e_q F^{\mu\nu}
<0|\bar{q}q|0>
\end{eqnarray}
Moreover, the $\theta$ dependence of $<0|\bar{q}\Gamma
q|0>_{\theta}$ matrix elements may be traced by considering the
anomalous axial current \cite{PR}:

\begin{equation}
m_q <0|\bar{q}\Gamma q|0>_{\theta}= i m_*\theta <0|\bar{q}\Gamma
q|0> +  O(m_q^2 )
\end{equation}
where the correction $O(m_q^2 )$ is negligible since $m_\eta >>
m_\pi $.

Putting altogether the above ingredients and after a
straightforward calculation \cite{PR1}, the following expression
of $\Pi^{(1)}(q^2,F^{\mu\nu})$ for the neutron is derived:

\begin{eqnarray}
\Pi(-q^2)&=&-{\bar{\theta}m_* \over
{64\pi^2}}<0|\bar{q}q|0>\{\tilde{F}\sigma,\hat
q\}[\chi(\beta+1)^2(4e_d-e_u) \ln({\Lambda^2\over
-q^2})\nonumber\\ && -4(\beta-1)^2e_d(1+{1\over4}
(2\kappa+\xi))(\ln({-q^2\over \mu_{IR}^2})-1){1\over
-q^2}\nonumber\\ &&-{\xi\over
2}((4\beta^2-4\beta+2)e_d+(3\beta^2+2\beta+1)e_u){1\over
-q^2}...],
\end{eqnarray}
with $\hat q=q_\mu\gamma^\mu$. \\
The QCD expression (11) is confronted to the phenomenological
parameterization $\Pi^{Phen}$$(-q^2)$ written in terms of the
Neutron hadronic properties. The latter is given by:

\begin{equation}
\Pi^{Phen}(-q^2)=\{\tilde{F}\sigma,\hat q\}
({\lambda^2d_nm_n\over(q^2-m_n^2)^2} +{A\over (q^2-m_n^2)}+...),
\end{equation}
where $m_n$ is the neutron mass, $e_q$ is the quark charge. the
parameters A and $\lambda^2$, which  originate from the
phenomenological side of the sum rule, represent respectively a
constant of dimension 2 and the neutron coupling constant to the
interpolating current $\eta(x)$. This coupling is defined via a
spinor $v$ as $<0|\eta(x)|n>=\lambda v e^{\alpha \gamma_5}$.

In the framework of QCD sum rules, the correlators at finite
temperature are expressed in terms of the thermal Gibbs average of
 Wilson operator expansion \cite{BS,M}. At relatively low
temperature, where the system can be regarded as a non interacting
gas of bosons, the thermal dependence of the vacuum condensates
can be written as :
\begin{equation}
<O^i>_T=<O^i>+\int{d^3p\over
2\epsilon(2\pi)^3}<\pi(p)|O^i|\pi(p)>n_B({\epsilon\over T})
\end{equation}
where $\epsilon=\sqrt{p^2+m^2_\pi}$, $n_B={1\over{e^x-1}}$ is the
Bose-Einstein distribution and $<O^i>$ is the standard vacuum
condensate (i.e. at T=0). In this approximation, we only kept the
pion contributions, since in the low temperature region, the
effects of heavier resonances $(\Gamma= K, \eta,.. etc)$ are
dumped by their distibution functions $\sim e^{- m_\Gamma \over
T}$ \cite{K}. To compute the pion matrix elements, we apply the
soft pion theorem given by:
\begin{equation}
<\pi(p)|O^i|\pi(p)>=-{1\over f^2_\pi}<0|[Q^a_5,[Q^a_5,O^i]]|0>+
O({m^2_\pi \over \Lambda^2}),
\end{equation}
where $ \Lambda$ is a hadron scale and $Q^a_5$ is the isovector
axial charge defined by:
\begin{equation}
Q^a_5=\int d^3x \bar{q}(x)\gamma_0\gamma_5{\tau^a\over2}q(x).
\end{equation}
Direct application of the above formula to the quark  and gluon
condensates shows the following features \cite{GL,K}:\\
(i) Only $<\bar{q}q>$ is sensitive to temperature. Its behaviour
at finite T is given by:
\begin{equation}
<\bar{q}q>_T\simeq (1-{\varphi(T)\over8})<\bar{q}q>,
\end{equation}
where $\varphi(T)={T^2\over f^2_\pi}B({m_\pi\over T})$, $B(z)=
{6\over\pi^2}\int_z^\infty dy {\sqrt{y^2-z^2}\over{e^y-1}}$ and
$f_\pi$ is the pion decay constant ($f_\pi\simeq 93 MeV$). The
variation with temperature of the quark condensate $<\bar{q}q>_T$
results in two different asymptotic evolutions, namely: \\
$$<\bar{q}q>_T\simeq (1-{T^2\over {8f^2_\pi}})<\bar{q}q>$$ \quad for
 ${m_\pi\over T}\ll 1$,
 $$<\bar{q}q>_T\simeq (1-{\sqrt{\pi m_pi \over {2T}}T^2\over
{8f^2_\pi}}e^{-m_\pi \over T})<\bar{q}q>$$ \quad for ${m_\pi\over
T}\gg 1$.\\
(ii) The gluon condensate is nearly constant at low
temperature and a T dependence occurs only at order $T^8$.

The determination of the ratio ${d_n \over \bar{\theta}}$ sum
rules at non zero temperature is now easily performed by applying
Borel operator to both parameterizations of the Neutron
correlation function shown in (11) and (12). Then finite
temperature effects are introduced via the procedure discussed
above. Finally, by invoking the quark-hadron duality, we deduce
the final sum rules of the $\bar { \theta}$ induced NEDM at
finite temperature:

\begin{equation}
{d_n\over \bar{\theta}}(T)=-{M^2m_* \over 16\pi^2}{1\over
\lambda_n^2(T)M_n(T)}(1-{\varphi(T)\over
8})<\bar{q}q>[4\chi(4e_u-e_d)-{\xi\over 2M^2}(4e_u+8e_d)]e^{M_n^2
\over M^2},
\end{equation}
where M represents the Borel parameter. \\
In order to get rid of the infrared divergence, the value of
$\beta$ has been set to 1 in (17). The Thermal evolution of the
coupling constant and the mass of the neutron were determined
from the thermal nucleon  sum rules \cite{K}.

Within the dilute pion gas approximation, Eletsky has shown that
the contribution induced the pion-nucleon scattering has to be
considered \cite{E}. It enters the nucleon sum rules through the
coupling constant $g_{\pi NN}$, whose values lie within the range
13.5-14.3 \cite{PROC}.
\\ \qquad Numerical analysis is performed with the following input
parameters: the Borel mass has been chosen within the values
$M^2=0.55-0.7GeV^2$ which correspond to the optimal range (Borel
window) in the $ d_n\over \bar{\theta}$ sum rule at $T=0$
\cite{PR1}. For the $\chi$ and $\xi$ susceptibilities we take
$\chi=-5.7\pm 0.6 GeV^{-2}$ \cite{BK} and $\xi=-0.74\pm 0.2$
\cite{KW}. As to the vacuum quark condensate appearing in (17),
we use its standard values \cite{SVZ}.

\section{Analysis and Conclusion}

We have established the relation between the NEDM and $
\bar{\theta} $ angle at non zero temperature from QCD sum rules.
We find that the behaviour of the ratio${ d_ n \over\bar{\theta}}$
 is connected to the thermal evolution  of the pion parameters
$f_\pi$, $m_\pi$ and of $g_{\pi NN}$.\\
 By analyzing the ratio  as a function of T in the region of validity of thermal sum-rules $[0,T_c]$, we learn that  $\mid{ d_ n \over \bar{\theta}}\mid$  decreases smoothly
with T (about 16$\%$ variation for temperature values up to 200
MeV) but survives at finite temperature. This means that either
the NEDM value decreases or $\bar{\theta}$ increases.
Consequently, for a fixed value of $\bar{\theta}$ the NEDM
decreases but it does not exhibit any critical behaviour.
Furthermore, if we start from a non vanishing $ \bar{\theta} $
value at $T=0$, it is not possible to remove it at finite
temperature. We also note that $ \mid{d_n\over \bar{\theta}}\mid$
grows as $M^2$ or $\chi$ susceptibility increases. It also grows
with quark condensate rising. However this ratio is insensitive
to both the $\xi$ susceptibility and the coupling constant
$g_{\pi NN}$. We notice that for high temperatures, the analysis
of $\mid{d_n\over \bar{\theta}}\mid=f({T\over T_c})$ exhibits a
brutal increase justified by the fact that for T beyond
 the critical value $T_c$, at which the chiral symmetry is restored,
the constants $f_\pi$ and $g_{\pi NN}$ become zero and
consequently the ratio $ {d_n\over \bar{\theta}}$ behaves as a
non vanishing constant. The large discrepancy between the values
of the ratio for $T<T_c$ and $T>T_c$ may originate from the other
contributions to the the spectral function which have been
neglected, such as the scattering process $ N+ \pi \to \Delta $.
These contributions, which are of the order $T^4$, are negligible
in the low temperature region but become substantial for $T\ge
T_c$. Moreover, this difference may also be due to the use of soft
pion approximation which is valid essentially for low $T$ ($T<
T_c$). Therefore it is clear from this qualitative analysis,
which is based on the soft pion approximation, that temperature
does not play a fundamental role in the suppression of the
undesired $\theta$-term and hence the broken CP symmetry is not
restored \cite{Chabab}. Indeed, some exact symmetries can be
broken by increasing temperature \cite{W,MS}. The symmetry non
restoration phenomenon, which means that a broken symmetry at T=0
remains broken even at high temperature, is essential for discrete
symmetries, CP symmetry in particular. Indeed, the symmetry non
restoration is a crucial ingredient in solving the domain wall
problem \cite{ZKO} and to create the baryon asymmetry  in the
early universe (BAU) \cite{S}.

This work is partially supported by  the convention de
cooperation between CNRST-Morocco/GRICES-Portugal 681.02/CNR, and
by the PROTARS III' grant D16/04.

\end{document}